\documentclass[epjST,final]{svjour}
\usepackage{graphics}
\usepackage{subfig}
\usepackage{url}
\begin{document}
\title{Differently Shaped Hard Body Colloids in Confinement: From passive to active particles}
%
%
\author{H.~H.~Wensink\inst{1} \and H.~L\"owen\inst{2} \and M.~Marechal\inst{3} \and A.~H\"artel\inst{4} \and R.~Wittkowski\inst{5} \and U.~Zimmermann\inst{2} \and A.~Kaiser\inst{2} \and A.~M.~Menzel\inst{2}}
\institute{
Laboratoire de Physique des Solides,
Universit\'e Paris-Sud and CNRS, 
91405 Orsay, France  
\and 
Institut f\"ur Theoretische Physik II: Weiche Materie, 
Heinrich-Heine-Universit\"at D\"usseldorf, 
40225 D\"usseldorf, Germany
\and
Institut f\"ur Theoretische Physik I,
Universit\"at Erlangen-N\"urnberg,
91058 Erlangen, Germany
\and
ITF, Utrecht University,
MG 305, Leuvenlaan 4,
3584 CE Utrecht, The Netherlands 
\and
SUPA, School of Physics and Astronomy, 
University of Edinburgh,
Edinburgh EH9 3JZ, United Kingdom
}

\abstract{ 
We review recent progress in the theoretical description  
of anisotropic hard colloidal particles. The shapes considered range from rods and dumbbells to
rounded cubes,  polyhedra to biaxial particles with arbitrary shape.
Our focus is on both static and dynamical 
density functional theory and on computer simulations.  We describe recent results for the
structure, dynamics  and phase behaviour in the bulk and in various confining geometries, e.g. established by two parallel walls which 
reduce the dimensionality of the system to two dimensions. We also include recent 
theoretical modelling for active particles, which are autonomously driven 
by some intrinsic motor, and highlight their fascinating nonequilibrium dynamics and 
collective behaviour. 
} 
\maketitle
\section{Introduction}
\label{intro}
Nowadays anisotropic colloidal particles can be prepared in a controlled way and almost any shape
can be realized \cite{ivlev2012complex,Wittkowski2012MolPhys,glotzer2007}. In their simplest form, the particles 
are sterically stabilized so that their 
interactions can be described by excluded volume alone. 
This implies that temperature scales out (the thermal energy $k_{B}T$ trivially 
sets the energy scale)
such that in  the bulk the density (or the packing fraction)
 is the only relevant thermodynamic parameter. In the absence of external fields, the dynamics of
a colloidal particle is Brownian due to random forces exerted by the solvent molecules. 
The additional degree of anisotropy has given a boost to 
the development of new statistical
theories for anisotropic particles and extended methods for computer simulations.
For example, the traditional theory of Brownian motion has 
been widely applied to spherical and rod-like particles 
with a rotational symmetry \cite{dhont2003introduction,Lowen_1994_PRE}. Case studies 
of particles with an arbitrary shape have not yet been fully considered in theory and simulation, although the theoretical principles date back to 1934 when Perrin \cite{Perrin1934} developed the basic ideas. 
Density functional theory is typically formulated for systems of spherical particles whereas generalized theories appropriate for systems with orientational degrees of freedom are scarce.

A system of hard nonspherical particles can be exposed to an 
external confinement, which leads to even more complex behaviour, see the complexity diagram in 
\cite{Lowen2001}.  
The simplest types of confinement are an external planar hard wall 
(see e.g.\  \cite{Dullens,Sandomirski}), 
two parallel hard plates (``slit geometry'') \cite{Neser_PRL_1997},
or a templated wall \cite{Hoogenboom1}. Other options are
periodic modulations induced by a laser field (so called optical gratings) 
\cite{JenkinsEgelhaaf,EPJEST_19}, which are found to induce
highly non-trivial phase behaviour for simple spheres 
\cite{Wensink2007JPCM,Wensink2007JCP,Virgiliis2007,Vink2007,Brader2007,Royall2007_PRL,van2006isotropic,Vink}.
The extreme limit of a slit geometry is a complete reduction in dimensionality 
down to two spatial dimensions. The same reduction can be achieved if 
colloids are confined to interfaces \cite{EPJEST_15,EPJEST_24}.
In general, confinement brings in  additional system parameters which require a more generic statistical 
theory to describe inhomogeneous systems.

In contrast to passive particles,
self-propelled particles dissipate energy and exhibit autonomous, directed motion.
 This leads to intrinsic nonequilibrium behaviour
\cite{Romanczuk2012,Cates2012,Marchetti2012}. 
The self propulsion specifies a direction in orientational phase space, which renders active 
particles and their interactions inherently anisotropic, even though their shape might be spherical. 
Several varieties of self-propelled colloidal particles have been 
prepared in which external fields  play a leading role in the self-propagation mechanism. 
Prominent examples include catalytically driven Janus particles \cite{erbe2008various}
and thermally driven colloids in a phase-separating solvent 
\cite{Bechinger,KuemmelPRL2013,Buttinoni2013_PRL}. Last but not least,  autonomously moving bacteria (such as {\em Bacillus subtilis}) exhibit characteristics
similar to self-propelled colloidal particles \cite{Wensink_PNAS}.

In this paper, we review recent progress in the theoretical description  
of anisotropic hard colloidal particles. The various shapes considered range from simple rods and dumbbells to
rounded cubes,  polyhedra and to particles with arbitrary biaxiality. 
Here we mainly use static and dynamical density functional theory and we test the theory against computer simulations.  
We describe recent results for the
structure, dynamics  and phase behaviour 
in bulk and in various confining geometries. We also include results from recent 
theoretical modelling of ensembles of active particles 
and highlight the main features of their collective behaviour. 

The review is organized as follows: in sec.~\ref{dft} we present a general overview of results from density functional theory, 
both static and dynamical, for anisotropic particles before discussing some explicit results 
obtained from fundamental-measure theory for rods \cite{Haertel,Cremer}, dumbbells 
\cite{dumbbells}, rounded cubes \cite{parallelcubes}, and regular  polyhedra \cite{polyhedra} in sec.~\ref{passive_colloids_statics}.
In sec.~\ref{passive_colloids_dynamics}, we turn to Brownian dynamics of anisotropic particles, both in equilibrium and  nonequilibrium situations.
Section~\ref{active_colloids} is devoted to single-particle and collective behaviour of self-propelled colloids and we conclude in sec.~\ref{conclusions}.
\section{Density functional theory}
\label{dft}
\subsection{Static density functional theory: Fundamental-measure theory}
\label{sec:static_dft}
Classical density functional theory of freezing 
\cite{Evans1979,Singh1991,Loewen_1994,Tarazona_review,Wu}
provides an ideal framework to describe nonuniform fluids. Confinement can be studied in a very elegant way, because an external potential enters very naturally in the theoretical framework.
One of the most successful  
approximations for the hard sphere functional is based on the fundamental-measure theory (FMT) of Rosenfeld \cite{Rosenfeld89} (see \cite{Roth2010} for a review), which is purely geometry-based. 
This theory has been extended to colloid-polymer mixtures \cite{Brader}(see \cite{Vink} for an application). More importantly, it has been generalized
towards hard particles with arbitrary shape (for related work see \cite{Chamoux1996_JCP,Korden2012_PRE}) by Hansen-Goos and Mecke \cite{MeckePRL,HansenGoosJPCM}
who have extended the FMT framework by including the orientational degrees of 
freedom of biaxial particles \cite{Wittkowski2012MolPhys}. In the extended fundamental-measure theory (henceforth referred to as E-FMT) the one-body density
depends on the centre-of-mass position and orientation while particle interactions are assumed to be hard.

Although the E-FMT  provides a general framework to study a vast repertoire of particle shapes, only the case of hard spherocylinders in the fluid phase was considered in the original paper \cite{MeckePRL}.
Recent efforts have focussed on testing the 
applicability of  E-FMT  for other particle shapes such as
dumbbells \cite{dumbbells}, rounded cubes \cite{parallelcubes} and regular  polyhedra
\cite{polyhedra}. We summarize these situations in the following sections.
\subsection{Dynamical density functional theory for biaxial particles}
\label{sec:dynamic_dft}
On large timescales colloidal particles exhibit completely overdamped Brownian dynamics due to solvent friction.
For spherical Brownian particles, dynamical density functional theory (DDFT) was derived by
either starting from the Langevin equations  \cite{Marconi2000}, the Smoluchowski equation 
\cite{Archer04}, or the Mori-Zwanzig projection operator technique \cite{Espanol2009}(see \cite{Advances} for a recent review). 
Subsequently, a DDFT for uniaxial anisometric particles \cite{Rex} has been set up based on
the Smoluchowski-based approach proposed in \cite{Archer04}. The case of uniaxial particles involves three friction (or mobility) coefficients, namely the translational 
friction parallel and perpendicular to the main particle orientation and the rotational friction.

More recently, DDFT has been extended to rigid biaxial particles with an arbitrary shape \cite{Wittkowski2012MolPhys}.
In these situations the friction matrix is much more complicated due to various nontrivial translational-rotational couplings. This is important for, e.g. screw-shaped particles where an imposed translational movement induces rotational motion and vice versa. 
\subsection{Phase-field-crystal models for liquid crystals}
\label{sec:pfc}
In density functional theory (DFT), the one-particle density depends on the centre-of-mass position and on the orientation.
For uniaxial particles, the orientational dependence can be expanded into spherical harmonics such that the density is parametrized with just position-dependent order-parameter fields such as the mean 
density, polarization, and  nematic order \cite{Advances,Wittkowski2010a}.
Using an order-parameter gradient expansion \cite{Beier,Lutsko,Elder2,Sven},  one can use DDFT as a starting point to derive simpler theories for the dynamics 
of liquid crystals. For spheres, these class of theories are usually referred to as phase-field-crystal (PFC)
models as originally introduced in 2002 by Elder and coworkers \cite{Elder1}(see \cite{Advances} for a review). PFC theory is capable of predicting a vast range of phase transitions to positionally ordered structures  commonly observed in hard and soft condensed matter.
A generalized PFC model for systems with orientational degrees 
of freedom has been recently developed in both two \cite{LowenPFCLC} and three 
\cite{Wittkowski2010a} spatial dimensions. The simplified theory is still microscopic 
in the sense that the PFC coupling parameters can be related to various microscopic properties of the system such as the pair potential and pair correlation function.
As such the extended PFC  model provides an approximate yet microscopically founded framework for 
the calculation of phase diagrams of liquid crystals \cite{Achim}. Moreover, it can be used as a useful starting point for addressing dynamical processes  of anisotropic particles since the PFC models are technically less demanding than their DDFT counterparts.
\section{Passive colloids: statics}
\label{passive_colloids_statics}
\subsection{Hard spherocylinders}
\label{sec:hard_spherocylinders}
The phase diagram of spherocylinders has been computed by Bolhuis and Frenkel \cite{Bolhuis_frenkel}  
and features  stable isotropic, nematic, smectic A phases, as well as plastic and orientationally-ordered
 $AAA$ or $ABC$ stacked crystals.
The work of Bolhuis and Frenkel \cite{Bolhuis_frenkel}
 thus provides excellent benchmark data to test density functional theories for rodlike colloids \cite{Graf}.
The recently developed E-FMT 
has been applied to inhomogeneous isotropic and nematic phases \cite{Haertel} but the 
computation of the full phase diagram is still hampered by the enormous numerical effort 
required to evaluate the functional for structured phases such as smectic A and plastic crystal.
Preliminary calculations, however, seem to indicate  that there is no stable smectic A phase \cite{RRoth}.
This result clearly contradicts observations from simulation and experiment and thus prompts a careful study of the numerical implementation of E-FMT.
It would also be interesting to address the plastic-crystalline state. In particular,  one could explore the 
nontrivial topology of the director field inside a unit cell which has
recently been unveiled using computer simulations \cite{Cremer}.

The numerical burden can be significantly reduced by applying E-FMT to spherocylinders 
in strong aligning fields which only couple to the orientational degrees of freedom \cite{Graf}. Numerical results for the 
isotropic and nematic fluids of spherocylinders and their coexistence are presented in \cite{Haertel}.
\subsection{Hard dumbbells}
\label{sec:hard_dumbbells}
A hard dumbbell consists of two fused hard spheres of diameter $\sigma$ whose centres
are separated by a distance $L < \sigma$. Note that this produces a nonconvex shape. Different synthesis methods have
been successfully applied to prepare colloidal dumbbells 
\cite{johnson2005synthesis,Mock2006,Lee2008}. Samples of colloidal dumbbells
have been used to study  bulk crystallization \cite{Mock2007} and
quasi two-dimensional degenerate crystals 
\cite{Lee2008,Gerbode2008,gerbode2010glassy}.
Moreover, the bulk phase behaviour of hard dumbbells is well-known from simulations
\cite{vega1992solid,Vega1992stability,Vega1997,Vega2004,marechal2008stability}.

\begin{figure}
  \centering 
  \resizebox{1.0\columnwidth}{!}{\includegraphics{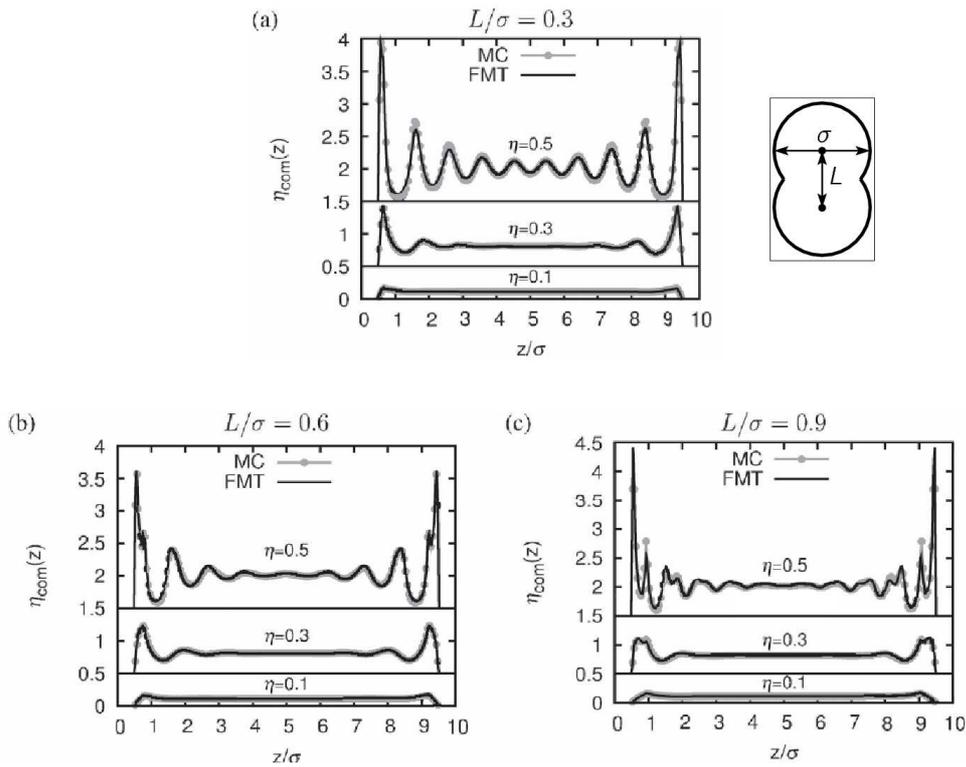}}
  \caption{The centre-of-mass density profile $\eta_\mathrm{com}(z)$, 
in units of the volume of a dumbbell,
as a function of the height $z$ between two parallel walls for dumbbells of diameter $\sigma$ 
and three different values of $L/\sigma$ at bulk volume fractions $\eta$. Monte-Carlo simulation results are denoted 
by the gray dots and
the FMT results  by the black line. From \cite{dumbbells}.}
  \label{fig:1}   
  \end{figure}

In the recent work \cite{dumbbells}, E-FMT was applied to an inhomogeneous fluid of
hard dumbbells. For various values of $L/\sigma$ and varied particle concentration, density profiles
for dumbbells in a slit-geometry were computed and found to be in good agreement with simulations 
for bulk packing fractions $\eta$ up to about $0.3$. An example is shown in Fig.~\ref{fig:1}.

DFT is also an ideal tool to
explore sedimentation profiles \cite{Biben}. Therefore, density profiles of dumbbells in a gravitational field were computed and compared to simulations \cite{Marechal2011}. 
 Again reasonable agreement was 
found between simulation and DFT. This shows 
that the E-FMT  \cite{MeckePRL,HansenGoosJPCM} gives an accurate prediction of the structure of hard 
particles with a
nonconvex shape at low to moderate particle density. Deviations between theory and simulation appear at 
higher packing fractions. These results suggest that the applicability  of E-FMT is limited to particle shapes that do not deviate too much from spherical particles.

\subsection{Hard rounded cubes}
\label{sec:hard_rounded_cubes}
\begin{figure}
\centering
\resizebox{0.65\columnwidth}{!}{\includegraphics{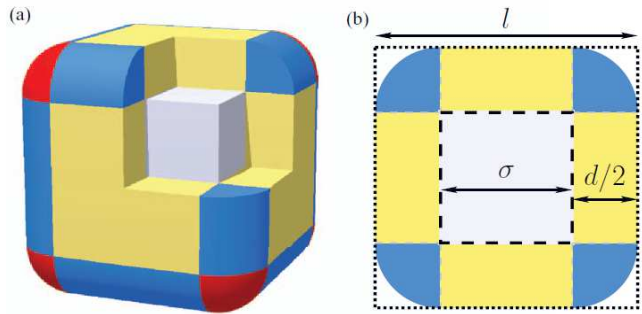}}
\caption{(a) A spherocube or rounded cube consists of a cube (lightest/gray)
surrounded by 6 square prisms (darker/yellow), 12 cylinder sections (still
darker/light blue), and 8 spherical sections (darkest/red). Some sections of
the outer objects have been removed to show the gray cube. (b) Cross section
of the spherocube showing the edge length $\sigma$, minimum radius of curvature
$d/2$ and the total width $\ell$. From \cite{parallelcubes}.}
\label{fig:2}   
\end{figure}

In the recent pioneering work of Rossi et al. \cite{Rossi2011},
 well-defined colloidal cubes were prepared and studied in real space by confocal
microscopy. These cubes typically possess rounded
edges, motivating a modelling by a hard ``spherocube'' with rounded edges of 
outer curvature diameter $d$ and an inner cubic part of edge length $\sigma$ (see Fig.~\ref{fig:2} for a sketch). 
Colloidal cubes can in principle be
oriented by external aligning fields \cite{HernandezNavarro2011,Kim2005},
 for instance by introducing
an inner core with two distinct nonparallel dipole
moments, each of which couples to a separate external field. Considering perfectly aligned cubes greatly facilitates the theoretical analysis,
since the orientational degrees of freedom of the cubes do not play a role.

\begin{figure}
\centering
\resizebox{1.0\columnwidth}{!}{\includegraphics{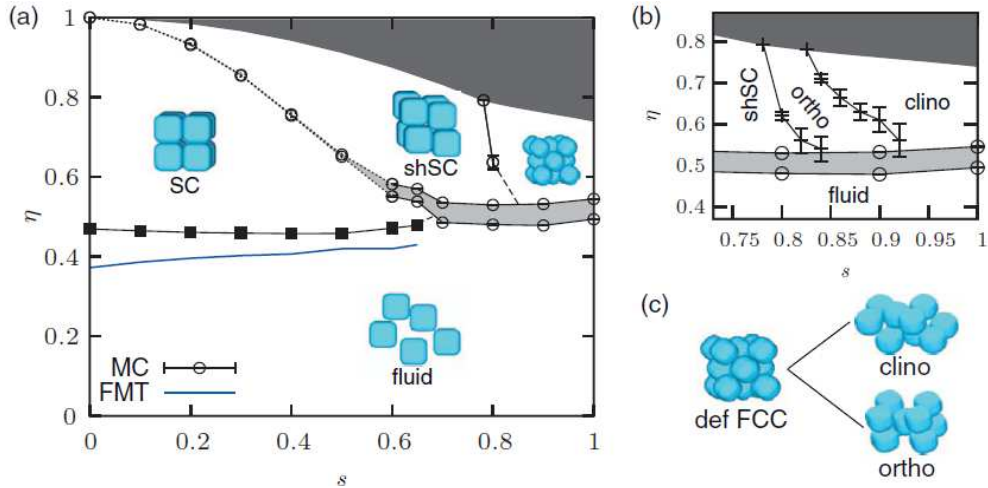}}
\caption{(a) Phase diagram of parallel rounded cubes in the $s$-$\eta$
representation, where $\eta$ is the packing fraction  and $s = d/\ell$
 is the rounding parameter (see Fig.~\ref{fig:2}). A cube corresponds to $s = 0$ and a sphere to $s = 1$. 
Shown are the areas of stability of the deformed
fcc phase of near spheres (def FCC), the sheared cubic crystal (shSC),
 the simple cubic crystal (SC), and the fluid phase in white. The forbidden region
above the close packing density is shown in dark gray and coexistence areas in lighter gray 
(coexistence lines are vertical). The filled symbols (MC simulations)
and the thick line (FMT) denote second-order phase transitions, 
while the empty symbols denote first-order phase transitions from simulations.
(b) Enlarged view of the large-$s$ region of the phase diagram: the def FCC phase 
is actually seen to have a body-centred orthorhombic variant (ortho) and a
base-centred monoclinic variant (clino), as depicted in (c). From \cite{parallelcubes}.}
\label{fig:3}   
\end{figure}

The bulk phase diagram of parallel hard spherocubes was recently derived from simulations
\cite{parallelcubes} and the results are  shown in Fig.~\ref{fig:3} as a function of the rounding parameter
$s = d/\ell$ and the cube packing fraction $\eta$. The special case $s = 0$
 of parallel cubes was studied earlier by Cuesta and coworkers 
\cite{Cuesta1996,cuesta1997fundamental,Cuesta1997dimensional,MartinezRaton1999} and 
a  second-order freezing transition into a simple cubic lattice was found. Second-order freezing is quite unusual in three dimensions and 
is intimately linked to the (cubic) anisotropy of the fluid phase along with the
properties of the simple cubic solid.
A second-order freezing transition occurs over a broad range of $s > 0$ (see Fig.~\ref{fig:3}). E-FMT calculations
 confirm the second-order character of the phase transition and give a transition packing fraction $\eta$
in good agreement with the simulation values, as evident from Fig.~\ref{fig:3}. 
Moreover, the simulations predict new crystal structures at
larger values of $s$ and high density corresponding to a sheared cubic lattice and a 
deformed face-centred-cubic (fcc) lattice which occurs in two variants: {\it ortho\/} and {\it clino}.
We emphasize that the present phase diagram is completely different from that of freely oriented 
hard cubes which exhibit a first-order freezing transition and
vacancy-stabilized crystalline order  \cite{Marechal_PNAS}.
The case of hard squares in two spatial dimensions has been studied
by Belli and coworkers \cite{Belli2012}. 

\subsection{Hard polyhedra}
\label{sec:hard_polyhedra}
The close-packed configuration of polyhedra, in particular tetrahedra 
\cite{Conway2006,Kallus2010,Chen2010,Jaoshvili2010}, has been intensely discussed as of late
\cite{Torquato2009,Graaf}. At intermediate densities novel states show 
up \cite{Torquato_review,Damasceno2012}, such as an emergent 
quasicrystalline state for hard tetrahedra \cite{haji2009disordered}.
Regarding theory for hard polyhedra there are some equation-of-state predictions 
(see \cite{polyhedra} for a discussion), but a microscopic theory for inhomogeneous 
systems has not been analyzed. 

In order to fill this gap, the E-FMT \cite{MeckePRL,HansenGoosJPCM}
has recently been applied to the layering of orientable hard polyhedric 
fluids near a hard wall \cite{polyhedra}. Density profiles for fluids 
of hard tetrahedra, cubes, octahedra and dodecahedra near a hard wall are shown in Fig.~\ref{fig:4}.
There is reasonable agreement with  simulation data, although the agreement deteriorates 
for low-order polyhedra (like tetrahedra). The fact that highly regular 
polyhedra, whose shapes closely resemble that of a sphere, can be accurately described by E-FMT  underlines the notion that the theory is 
optimized for (near-)spherical objects.
The density profiles plotted as a function of the centre-of-mass position, 
exhibit characteristic peaks which correspond to special configurations of a single  
polyhedron  close 
to a planar wall. The corresponding orientations are also depicted in Fig.~\ref{fig:4}.

\begin{figure}
  \centering
  \resizebox{0.6\columnwidth}{!}{\includegraphics{D3Fig4.eps}}
  \caption{Centre-of-mass density profiles $\eta(z)$ as a function of the height $z$ divided by
the edge length $\ell$ for one-component systems of (a) tetrahedra , (b) cubes, (c) octahedra, (d) dodecahedra and (e) icosahedra near a hard wall located at $z = 0$.
Fundamental measure density functional theory (DFT; dark, blue curves) is
compared to results from Monte-Carlo simulations (MC; light, orange curves).
The average volume fraction is set to $\eta=0.3$. (f) The density profile of spheres for the same
packing fraction $\eta=0.3$ is shown as a function of $z$ divided by the sphere diameter $\sigma$
for comparison. The heights $z$ for which a particle can just touch the wall with a vertex, an
edge and a face are indicated by the dashed lines;
also shown are polyhedra with the corresponding orientations. From \cite{polyhedra}.}
\label{fig:4} 
\end{figure}
\section{Passive colloids: dynamics}
\label{passive_colloids_dynamics}
\subsection{Brownian dynamics of biaxial particles}
\label{sec:bd_biaxial}
In three spatial dimensions, the Brownian
motion of colloidal particles with arbitrary shape involves a
hydrodynamic coupling between the translational and the rotational
degrees of freedom, which was described theoretically,
for example, by Brenner \cite{Brenner1965,Brenner1967}. 
In 2002, Fernandes and Garcia de
la Torre  \cite{Fernandes2002} proposed a corresponding Brownian dynamics
simulation algorithm for the motion of a passive rigid particle
with arbitrary shape. The underlying equations of motion were
generalized towards an imposed external flow field for the
surrounding fluid by Makino and Doi \cite{Makino2004}. Only recently, the full expression for the Langevin equations in Ito formalism including the drift term was derived in \cite{Wittkowski2012}, 
for the stochastically equivalent Smoluchowski formalism see \cite{Wittkowski2012MolPhys}. 
The predicted Brownian trajectories were experimentally confirmed \cite{Kraft2013_PRL}.
\subsection{Nonequilibrium dynamics in time-dependent external fields}
\label{sec:noneq_time_dependent_external_fields}
DDFT based on fundamental-measure theory for anisotropic particles has recently been employed in two cases concerning isotropic and nematic fluids of hard spherocylinders in time-dependent aligning fields.
First, the relaxational 
behaviour following an instantaneous switch in the direction of the aligning field was analyzed in \cite{Haertel}, which complements an earlier study considering a time-dependent translational 
confinement of soft rods \cite{Rex}. Second, an anisotropic  fluid was subject
to an aligning field rotating
at a  constant  angular speed. The nonequilibrium response of the system involves a number of novel nonequilibrium steady states
in the time-dependent orientational distribution \cite{Blaak}.
\section{Active colloids}
\label{active_colloids}
\subsection{Trajectories of a single self-propelled particle}
\label{sec:traj_single_self_propelled}
The simplest model for a single self-propelled (active) colloidal particle is a Brownian particle 
with an {\it internal\/} drive, i.e. with a propagation along an inner degree of 
freedom as modelled by an intrinsic ``force'' and ``torque''. In the absence of thermal noise 
the trajectories are closed circles in two spatial dimensions \cite{Teeffelen_circle,Zimmermann} but much more complex trajectories
arise in three spatial dimensions \cite{Wittkowski2012}. In particular the trajectories of orthotropic particles, which have no translational-orientational coupling in the hydrodynamic friction matrix, are helices. In general, the translational-rotational coupling of the hydrodynamic friction, however, leads to more complicated trajectories which have been classified in \cite{Wittkowski2012}.
The coupling to the thermal Brownian motion yields nontrivial behaviour
for the mean-square displacement, which can be worked out analytically 
for a constant intrinsic drive \cite{tenHagen} and  a uniaxial particle. For a
biaxial particle the problem becomes too complex and the average particle trajectories can only be assessed numerically \cite{Wittkowski2012}.
In two dimensions, the trajectory statistics were confirmed in experiments \cite{KuemmelPRL2013}.

Finally, we mention the motion of self-propelled particles in gradient fields of a chemical \cite{DietrichEPJE2010}.
In a poison field that weakens the drive, nontrivial scaling behaviour 
for the mean-square displacement has been obtained in \cite{Hoell} 
while the dynamics of a chemotactic predator-prey model
has been explored in \cite{Sengupta}.

\subsection{Collective properties of self-propelled rod-like particles}
\label{sec:collective_properties_self_propelled_rod}
The competition between the intrinsic self-propulsion and the interparticle interactions yields novel
emerging behaviour for dense systems of self-propelled colloids. Particle-resolved colloidal models consisting of self-propelled rods (SPR) interacting via a soft but strongly screened
Yukawa potential have become standard and can be simulated with and without
 Brownian noise \cite{Wensink_Lowen_2008}.  In fact, these types of segment models have been routinely used to describe fluids of  charged rods in thermal  equilibrium 
\cite{Kirchhoff_Lowen_Klein_1996,Lowen_1994}. Recently, these models have proven very useful in describing collective bacterial motion in {\em Bacillus subtilis}
and {\em Escherichia coli} which have an effective  aspect ratio of about 6 and 2, respectively \cite{Wensink_PNAS}. At higher aspect ratios,
self-propelled rods show a marked clustering in channel confinement, a phenomenon which is not encountered for passive rods in thermal equilibrium 
\cite{Wensink_Lowen_2008,Marceau_Gompper}. 

\begin{figure}
\centering
\subfloat[][]{\resizebox{0.6\columnwidth}{!}{\includegraphics{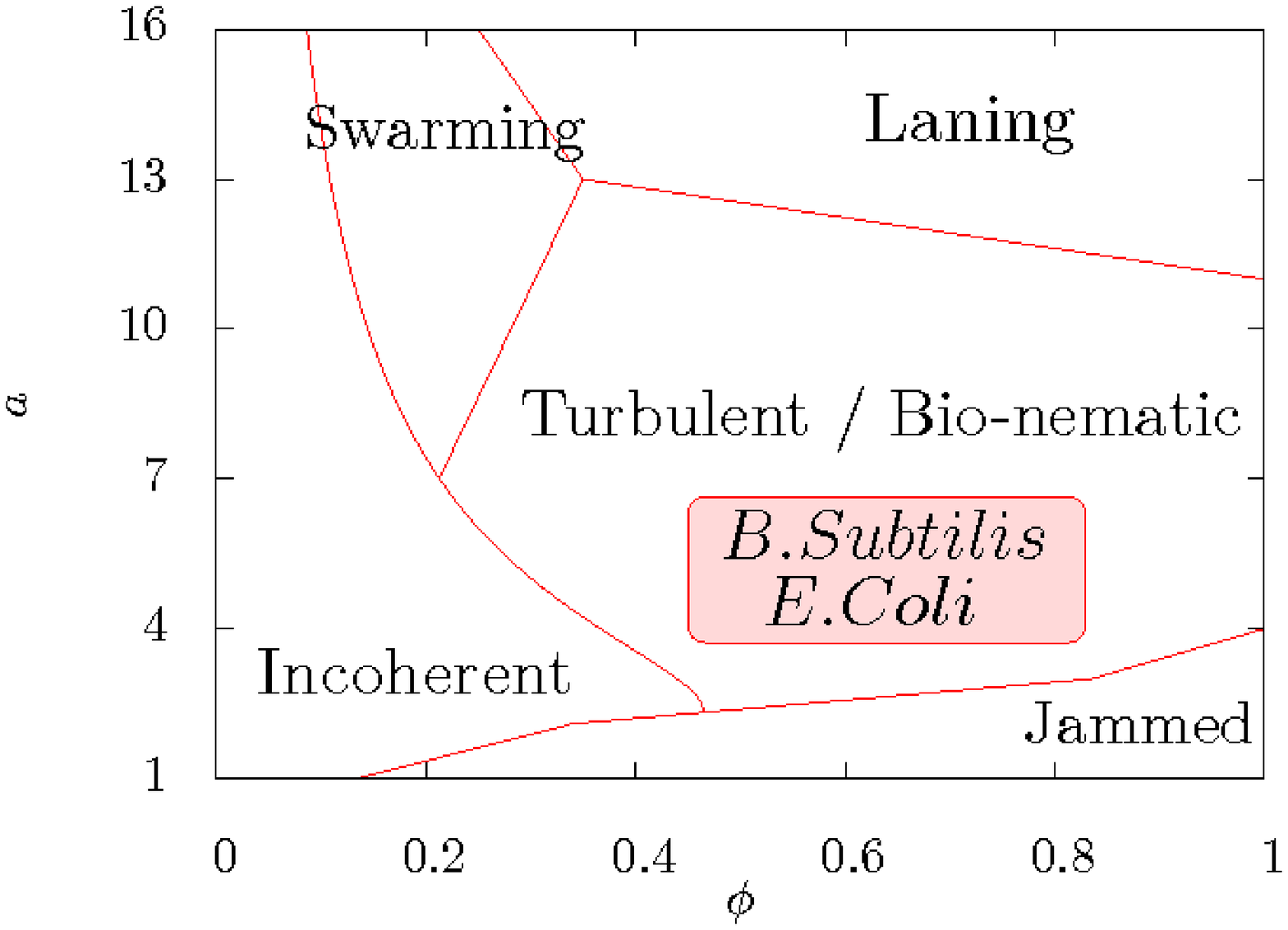}}}\\
\subfloat[][]{\resizebox{1.0\columnwidth}{!}{\includegraphics{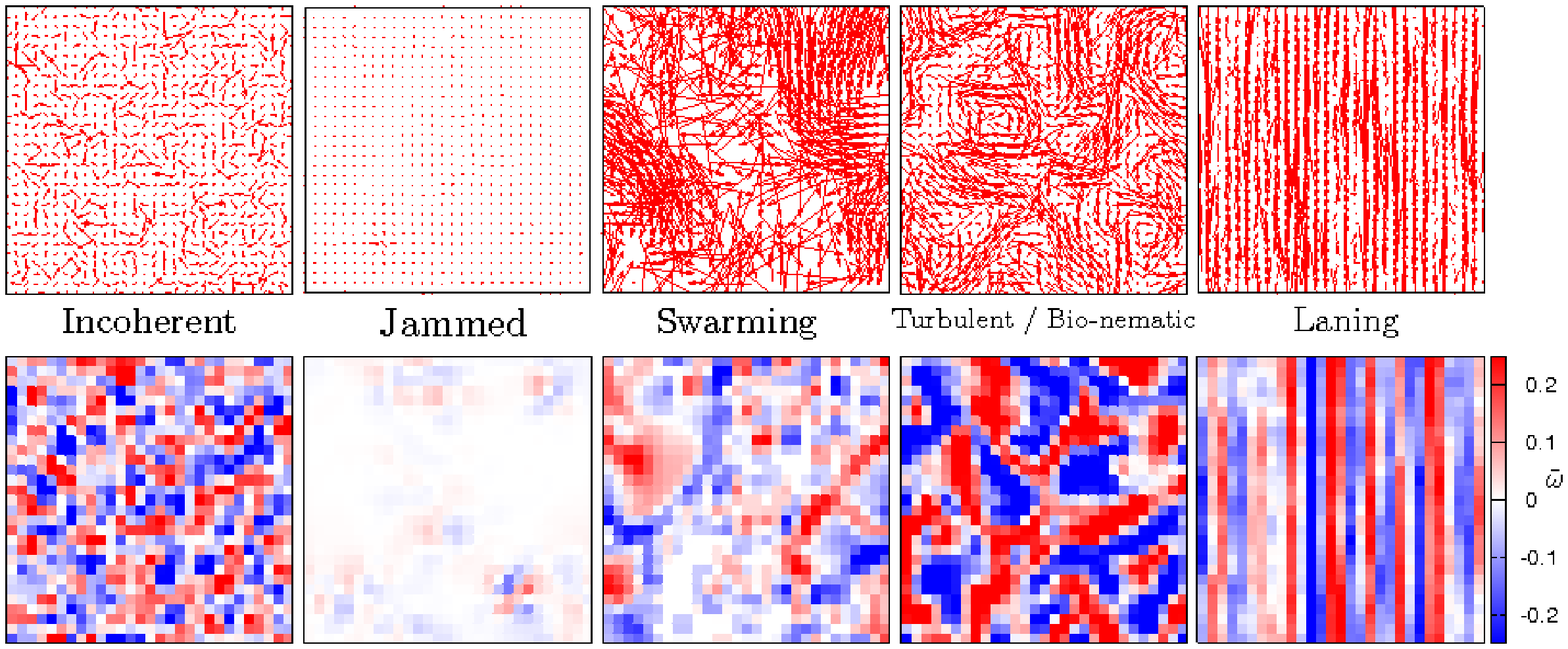}}}
\caption{(a) Schematic nonequilibrium phase diagram of self-propelled rods in two spatial
dimensions with a variable aspect ratio $a$ and effective filling fraction $\phi$. Values exceeding unity are, in principle,
possible due to the softness of the Yukawa interactions. The area relevant to self-motile
bacteria is highlighted in red. (b) A number of distinctly different dynamical states are
discernible as indicated by the coarse-grained maps of the velocity field  (upper 
panels) and the corresponding scalar vorticity field in the steady state (lower panels). From \cite{WensinkCODEF}.}
\label{fig:5}   
\end{figure}

In bulk situations the noiseless SPR model  exhibits a wealth of different 
steady states  \cite{WensinkCODEF,Wensink_PNAS}. These  are summarized in Fig.~\ref{fig:5} in a state diagram  spanned by the 
rod aspect ratio and the rod density. The reduced strength of the drive is fixed.
Using suitable order parameters, such as the vorticity of a coarse-grained velocity field, one can discern various emergent states: from a disordered one in weakly interacting systems to a  swarming state, a jammed state, and a laning state at higher density and aspect ratios.
Most notably there is an intermediate turbulent (or bio-nematic) state (Fig.~\ref{fig:5}) which is distinguished from the swarming state by a significant occurence of velocity swirls.
The topology of the phase diagram was confirmed by experiments on 
strongly confined {\em Bacillus subtilis} suspensions \cite{Wensink_PNAS}. The 
velocity correlations and energy spectrum in the turbulent flow do not 
follow the traditional Kolmogorov-Kraichnan scaling laws \cite{Wensink_PNAS}
which hints to a novel turbulent state in living fluids. In addition to linearly propagating rods, ``circle swimmers'' experience an additional active torque which may give rise to active self-assembly into vortex arrays \cite{Kaiser2013_PRE}. 

\begin{figure}
  \centering
  \subfloat[][]{\resizebox{0.6\columnwidth}{!}{\includegraphics{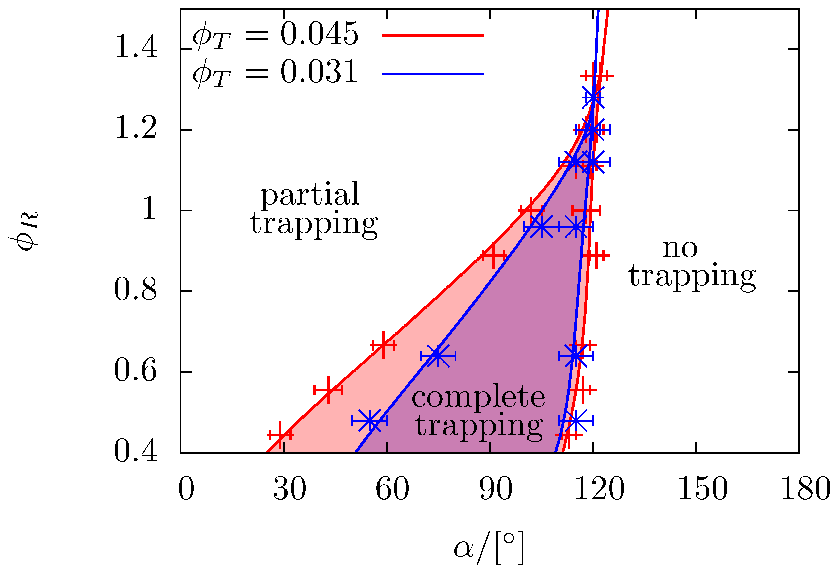}}}\\
  \subfloat[][]{\resizebox{1.0\columnwidth}{!}{\includegraphics{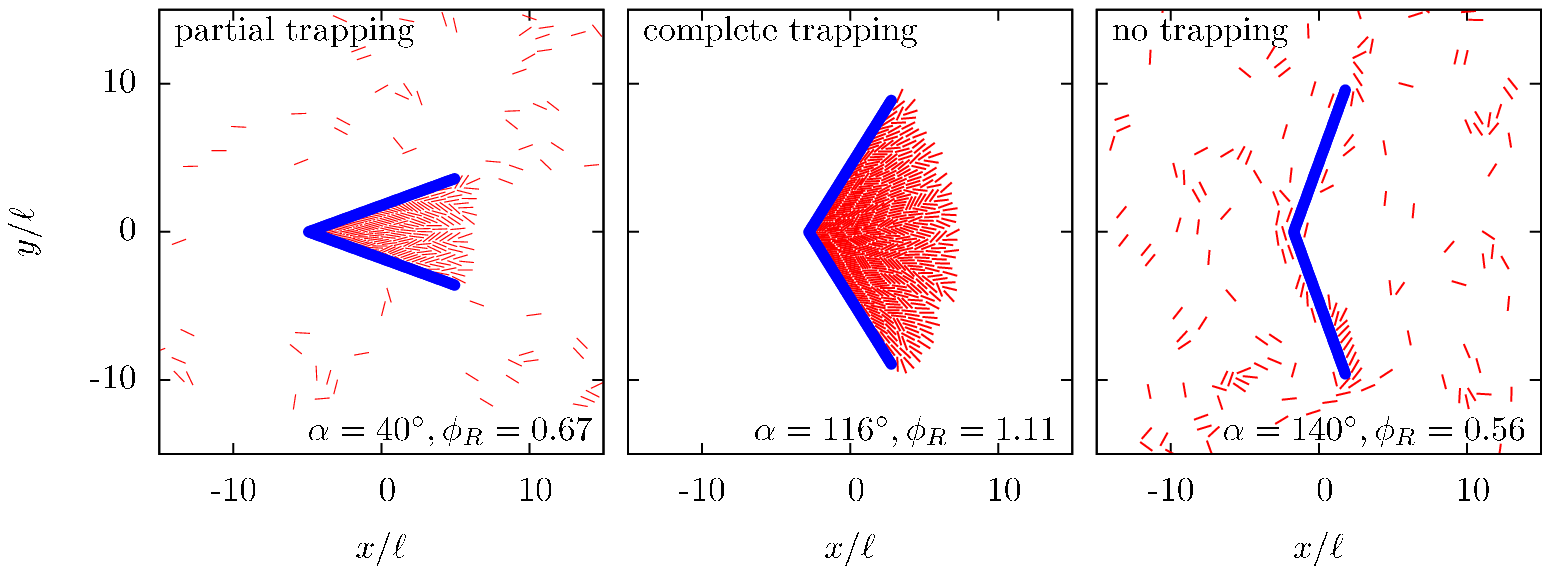}}}
  \caption{(a) Nonequilibrium phase diagram marking three different 
collective trapping states of SPRs  at a chevron boundary
with total length of 20 rod lengths $\ell$; no trapping at large apex angle $\alpha$, complete 
trapping at medium $\alpha$ and partial trapping at small $\alpha$ upon variation
of the reduced rod packing fraction $\phi_R = \phi/\phi_T$. Phase boundaries
 are shown for two different values of the area fraction $\phi_{T}$
occupied by the (periodically repeated) traps. The region of complete trapping is bounded 
by a triple point at larger rod concentration beyond which a smooth
transition from no trapping to partial trapping occurs. (b) Computer simulation snapshots  showing stationary configurations  of the three different states. From \cite{Kaiser}.}
  \label{fig:6}
\end{figure}

Next, the question of how to efficiently capture many self-propelled rods near fixed boundaries  was studied 
in \cite{Kaiser}. It was shown that a wedge-like trap with an appropriate opening angle constitutes
 an efficient trapping device for SPRs. The trapping efficiency has been classified in terms of a 
nonequilibrium phase diagram as a function of the wedge opening  angle,  the 
particle density and the trap density \cite{Kaiser}. This diagram is shown 
in Fig.~\ref{fig:6}  and highlights three possible final 
states of no trapping at all, partial trapping and
 complete trapping in which situation all particles  finally end up in the trap. We emphasise 
that the transition 
from partial to complete trapping is not a trivial consequence of the 
change in the (triangular) trap area. If the phase diagram were drawn at constant trap area, 
all three states would persist.  The trapping behaviour was also recently generalized towards a moving net \cite{Kaiser2013_PRE_2}. 
\subsection{Active crystals}
\label{sec:active_crystals}
``Active'' crystallization of self-propelled
 colloids was explored by means of particle-resolved computer simulations
\cite{Bialke}. Possibilities to explore ``active'' crystallization within microscopic PFC models and DDFT 
were reported in \cite{Wittkowski2012MolPhys}. It turned out that the bulk freezing transition in active systems occurs
 at much lower temperatures than in thermally equilibrated passive systems. Furthermore, at higher active drives or for spontaneous ordering of the self-propulsion directions, the crystal is predicted to propagate as a whole (``traveling crystal'') \cite{Menzel}. Clearly,  experiments on 
colloidal swimmers are called for to confirm these predictions.

\section{Conclusions}
\label{conclusions}
As documented by many applications, 
fundamental-measure density functional theory is a powerful predictive tool to describe the structure and dynamics of hard anisotropic particles. Dynamical 
density functional theory (DDFT)  provides a viable starting framework for developing 
a microscopic theory for active colloidal particles, which show fascinating collective 
behaviour like swarming, laning, turbulence and ``active'' crystallization.

We finalise by  outlining some unsolved problems which are interesting for future studies.
While there has been considerable progress in reformulating DFT for fluids in porous media \cite{SchmidtDFT,Wessels}, 
the corresponding dynamical extension has not been explored thus far. Future work should also consider 
mixtures of anisotropic particles (like rod-platelet mixtures, polydisperse systems) which are highly relevant to many colloidal systems in nature.   Mixing different shapes may lead to a wealth of new phase 
behaviours. Moreover, a DDFT with hydrodynamic interactions has been developed
for spheres \cite{RexPRL} but its extension to uniaxial and biaxial particles remains elusive. 
Finally, there is a keen interest to explore the dynamical response of anisometric particles in time-dependent confinement such as ``breathing'' traps \cite{Lowen2009}.

\begin{acknowledgement}
  This work was supported by the DFG within SFB TR6 (project D3).
\end{acknowledgement}

\bibliography{codef_refs}

\end{document}